\documentclass{IEEEtran}
\usepackage{cite}
\usepackage{hyperref}
\usepackage{amsmath,amssymb,amsfonts}
\usepackage{graphicx}
\usepackage{textcomp,nicefrac}
\usepackage{color, colortbl}
\definecolor{Gray}{gray}{0.9}
\definecolor{newcolor}{rgb}{.8,.349,.1}
\def\BibTeX{{\rm B\kern-.05em{\sc i\kern-.025em b}\kern-.08em
T\kern-.1667em\lower.7ex\hbox{E}\kern-.125emX}}
\usepackage{algorithm}
\usepackage{algpseudocode}

\begin{document}

\title{Unsupervised Learning of Multi-modal Affine Registration for PET/CT}

\author{Junyu Chen, Yihao Liu, Shuwen Wei, Aaron Carass, and Yong Du
\thanks{J. Chen and Y. Du are with the Department of Radiology and Radiological Science, Johns Hopkins University, MD, USA. (e-mail: jchen245@jhmi.edu). Y. Liu, S. Wei, and A. Carass are with the Department of Electrical and Computer Engineering, Johns Hopkins University, MD, USA.}\vspace{-5mm}}

\maketitle

\begin{abstract}
Affine registration plays a crucial role in PET/CT imaging, where aligning PET with CT images is challenging due to their respective functional and anatomical representations. Despite the significant promise shown by recent deep learning (DL)-based methods in various medical imaging applications, their application to multi-modal PET/CT affine registration remains relatively unexplored. This study investigates a DL-based approach for PET/CT affine registration. We introduce a novel method using Parzen windowing to approximate the correlation ratio, which acts as the image similarity measure for training DNNs in multi-modal registration. Additionally, we propose a multi-scale, instance-specific optimization scheme that iteratively refines the DNN-generated affine parameters across multiple image resolutions. Our method was evaluated against the widely used mutual information metric and a popular optimization-based technique from the ANTs package, using a large public FDG-PET/CT dataset with synthetic affine transformations. Our approach achieved a mean Dice Similarity Coefficient (DSC) of 0.870, outperforming the compared methods and demonstrating its effectiveness in multi-modal PET/CT image registration. The source code of this work is available at \url{https://bit.ly/3XTJrJh}.
\end{abstract}

\section{Introduction}
\label{sec:introduction}
\IEEEPARstart{A}{ffine} registration plays a crucial role in multi-modal imaging systems, including positron emission tomography (PET) and computed tomography (CT). The precise alignment of functional and anatomical images is essential for both image reconstruction in PET and subsequent theranostic procedures. Although deep learning (DL)-based registration methods have shown notable success in various medical imaging modalities and applications~\cite{chen2023survey}, their adoption in PET/CT has been limited, primarily relying on traditional pair-wise optimization-based methods. In this study, we make two key contributions: first, we explore the effectiveness of DL-based approaches for PET/CT image registration; second, we investigate an alternative to the commonly used mutual information (MI)---the correlation ratio (CR)~\cite{roche1998correlation}, which, while historically significant in multi-modal image registration, has been overlooked in recent times.
\section{Methods}
Let $m$ and $f$ represent the moving and fixed images, respectively. We deploy a deep neural network (DNN) to take in this image pair and predict a set of affine parameters $\alpha$. Our focus is on the three-dimensional (3D) scenario, which involves 12 total affine parameters: three each for translation, rotation, scaling, and shearing. These parameters are then applied in a warping function to align the moving image $m$ with the fixed image $f$. We adopt our previously developed network, \texttt{TransMorph} (TM)~\cite{chen2022transmorph}, which employs a Swin Transformer followed by multi-layer perceptrons to predict these parameters. The DNN is trained with the negative CR as the loss function (i.e., $\mathcal{L}(m\circ\alpha, f)$), detailed in section \ref{sec:CR}.

\vspace{-3mm}
\subsection{Multi-scale Instance-specific Optimization}
After the DNN predicts the affine parameters, these can serve as an initialization for the optimal transformation parameters. We then employ instance-specific optimization (ISO), iteratively refining these parameters for the corresponding image pair. This refinement is achieved by treating the affine parameters as model parameters and updating them through gradient descent. Although the DNN is trained on a dataset of images, it might not fully capture the unique characteristics of individual cases---a phenomenon known as the amortization gap~\cite{balakrishnan2019voxelmorph}. Therefore, ISO fine-tunes the affine parameters to the specific instance, enhancing the alignment accuracy. Given that there are only 12 parameters to adjust, this refinement process is exceptionally swift, especially on GPUs.

Here, we introduce a multi-scale ISO strategy that optimizes parameters across multiple image resolutions. The pseudocode for this process is detailed in Algorithm \ref{alg:1}. This approach involves five image scales. Throughout these scales, the affine parameters are progressively refined based on the image similarity measure (or the loss function) between the transformed moving image and the fixed image, leading to the optimized final transformation parameters.

\begin{algorithm}
\caption{Multi-scale instance-specific optimization}\label{alg:1}
\begin{algorithmic}
\State \textbf{Data:} $m$ and $f$; \textbf{Result:} $\alpha$
\State $\alpha=\text{DNN}(m,f)$ \Comment{DNN estimates affine parameters}
\State $scales=[16,8,4,2,1]$ \Comment{Downsampling factors}
\State $iters=[100,100,120,140,160]$ \Comment{Num. of iterations}
\For{$idx$ \textbf{in} \texttt{enumerate}($scales$)}
    \State $\hat{m}=\text{\texttt{downsample}}(m,scales[idx])$ 
    \State $\hat{f}=\text{\texttt{downsample}}(f,scales[idx])$
    \For{$iter$ \textbf{in} \texttt{enumerate}($iters$[$idx$])}
        \State $\ell=\text{\texttt{ImgSim}}(\hat{m}\circ\alpha,\hat{f})$
        \State $\alpha\gets\ell$ \Comment{Gradient descent to update $\alpha$}
    \EndFor
\EndFor
\end{algorithmic}
\end{algorithm}
\begin{figure*}[ht]
\centering
\includegraphics[width=0.98\textwidth]{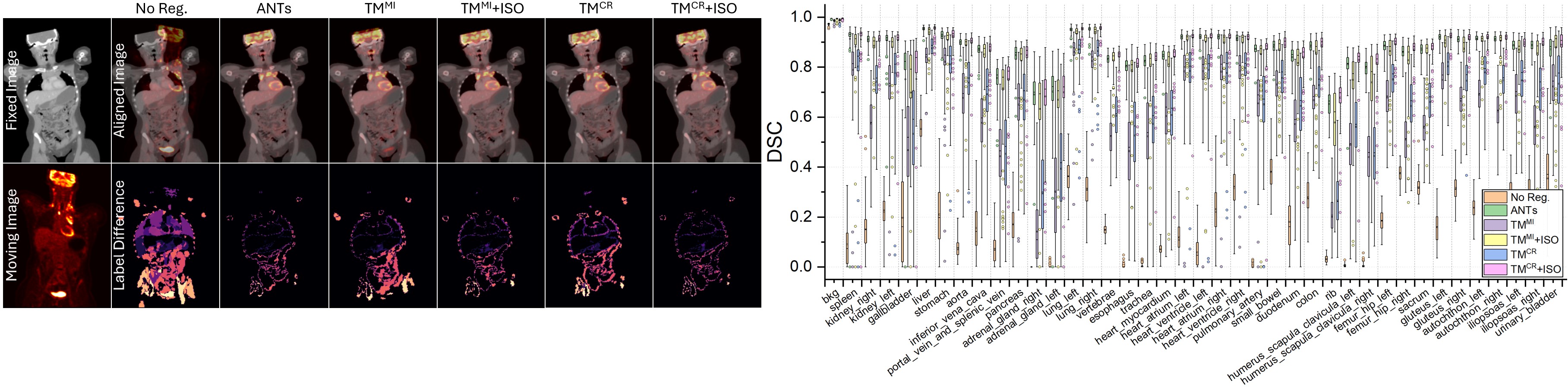}
\vspace{-3mm}
\caption{Qualitative and quantitative analysis comparing different affine registration methods, including scenarios without registration.}\label{fig:results}
\vspace{-5mm}
\end{figure*}
\vspace{-3mm}
\subsection{Differentiable Correlation Ratio via Parzen Windowing}
\label{sec:CR}
The CR for image registration was initially introduced in \cite{roche1998correlation}. It operates on the principle that the intensity values in $Y$ can be predicted using values from image $X$, where $X$ and $Y$ may belong to the same or different modalities. The CR is calculated as the ratio of the variance of the conditional expectation of $Y$ given $X$ to the variance of $Y$, and is mathematically expressed as $\eta(Y|X) = \frac{Var[E(Y|X)]}{Var(Y)}$.
In the original method, $\eta$ is implemented using a discrete method that involves iterating through discrete image intensity values, as detailed in~\cite{roche1998correlation}. Yet, such an implementation cannot be used for DNN training due to its non-differentiable nature. Instead, we employ a Parzen windowing method with Gaussian kernels to approximate the conditional expectation, similar to the approach used for MI in \cite{guo2019multi}. The conditional expectation given a particular intensity bin is approximated as:
\begin{equation}
    \Bar{y}_k = \frac{\sum_{i}\omega_{ik}y_i}{\sum_{i}\omega_{ik}},
\end{equation}
where $i$ denotes the $i$-th voxel location and $k$ represents the $k$-th intensity bin. Here, $\omega_{ik}$ is a Gaussian kernel, given by:
\begin{equation}
    \omega_{ik}=\frac{1}{h\sqrt{2\pi}}\exp{\Big(-\frac{(x_i-\text{bin}_k)^2}{2h^2}\Big)},
\end{equation}
where $\text{bin}_k$ is the center of the $k$-th intensity bin, and $h$ is the bandwidth of the Gaussian kernel. The variance of the conditional expectation is calculated as: 
\begin{equation}
    \sigma^2_{cond}=\sum_kn_k(\Bar{y}_k-\Bar{y})^2,
\end{equation}
where $n_k=\frac{\sum_iw_{ik}}{\sum_i\sum_kw_{ik}}$ and $\Bar{y}=\frac{1}{\Omega}\sum_iy_i$, with $\Omega$ representing the total number of voxels. The differentiable CR is then defined as $\eta(Y|X) = \frac{\sigma^2_{cond}}{\sigma^2}$, where $\sigma^2$ is simply the variance of image $Y$. Given that the CR is naturally asymmetric, where $\eta(Y|X)\neq\eta(X|Y)$, we address this by using a symmetric loss function $\mathcal{L}(X, Y)=-\frac{1}{2}\Big(\mathcal{\eta}(Y|X)+\mathcal{\eta}(X|Y)\Big)$ as the final form for our loss function. This loss is further extended to a local patch-based version, as similarly done in~\cite{guo2019multi}. This modified loss, $\mathcal{L}_{patch}$, is computed by averaging the negative CR across all local patches.

\vspace{-2mm}
\subsection{Dataset and Implementation Details}
Our evaluation of the proposed method was conducted using the AutoPET dataset~\cite{gatidis2022whole}, which comprises 896 FDG-PET/CT scans. We adopted a dataset division ratio of 7:1:2, resulting in 628, 90, and 178 images for training, validation, and testing, respectively. All images were resampled to the same voxel dimension of $2.8\times2.8\times3.8\ mm^3$. The CT images were processed through an automated segmentation algorithm, TotalSegmentator, which provided 40 anatomical label maps for method evaluation using Dice similarity coefficient (DSC).  For the registration task, random affine transformations were applied separately to the paired PET and CT images of the same subject, with the objective to align the PET image with the corresponding CT image. We compared our CR-based loss against the widely used MI, using the same DNN. The DNN was first trained using global versions of the loss functions, followed by local patch-based versions for ISO. As a baseline, a traditional optimization-based method employing MI from the ANTs package~\footnote{\url{https://github.com/ANTsX/ANTs}} was compared.
\begin{table}[h]
\centering
\caption{Summary quantitative results on the test set.}\label{tab:quant_res}
\begin{tabular}{ |c|c|c|c|} 
 \hline
     & No Reg. & ANTs & TM$^{MI}$ \\
 \hline
 \rowcolor{Gray}
 DSC & 0.202$\pm$0.054& 0.867$\pm$0.037 & 0.629$\pm$0.111 \\
  \hline
     & TM$^{MI}$+ISO & TM$^{CR}$ & TM$^{CR}$+ISO
\\ 
 \hline
 \rowcolor{Gray}
 DSC & 0.843$\pm$0.087 & 0.684$\pm$0.119&0.870$\pm$0.059
 \\ 
 \hline
\end{tabular}
\vspace{-3mm}
\end{table}

\section{Results and Conclusion}
Figure \ref{fig:results} presents both qualitative and quantitative comparisons among various methods. Results indicate that the conventional optimization-based approach using MI performs well, even surpassing DL methods that do not incorporate ISO. However, the introduction of multi-scale ISO significantly enhances alignment accuracy. As depicted in the box plot, the DNN trained using CR alongside ISO excels over all other methods in terms of performance across different organs, a finding corroborated by Table \ref{tab:quant_res}.

In this study, we introduced an unsupervised DL-based affine registration technique for PET-CT imaging. This involves a novel implementation of differentiable CR combined with a multi-scale ISO framework. The proposed method has been validated on a public dataset and demonstrates its efficacy in multi-modal affine registration.


\bibliographystyle{IEEEtran}
\bibliography{reference}

\end{document}